\documentclass[floatfix,aps,prd,showpacs,amsmath,nofootinbib,preprintnumbers,
twocolumn]
{revtex4}
\usepackage{graphicx}
\usepackage{colordvi}
    \def\be{\begin{equation}}
    \def\ee{\end{equation}}
    \def\ba{\begin{eqnarray}}
    \def\ea{\end{eqnarray}}
\begin{document}
% \documentclass[twocolumn,showpacs]{article}
%\documentstyle[aps,prd]{revtex4}
%\documentstyle{article}
%\usepackage{graphicx}
%\tightenlines
%\renewcommand{\thefootnote}{\fnsymbol{footnote}}
%\def\be{\begin{equation}}
%\def\ee{\end{equation}}
%\def\ba{\begin{eqnarray}}
%\def\ea{\end{eqnarray}}
%\draft
%\begin{document}

\title{Effect of accretion on primordial black holes in Brans-Dicke theory}

%\author{B. Nayak$^{1}$, A. S. Majumdar$^2$ and L. P. Singh$^{\dag{1}}$}

\author{B. Nayak} 
\altaffiliation{bibeka@iopb.res.in}
\affiliation{Department of Physics, Utkal University, Vanivihar,
Bhubaneswar 751004, India}

\author{A. S. Majumdar} 
\altaffiliation{archan@bose.res.in}
\affiliation{S. N. Bose National Centre for Basic Sciences, Salt Lake, 
Kolkata 700098, India}

\author{L. P. Singh} 
\altaffiliation{lambodar\_uu@yahoo.co.in}
\affiliation{Department of Physics, Utkal University, Vanivihar,
Bhubaneswar 751004, India}

%%%%%%%%%%%%%%%%%%%%%%%%%%%%%%%%%%%%%%%%%%%%%%
\begin{abstract}
We consider the effect of accretion of radiation in the early universe
on primordial black holes in Brans-Dicke theory. The rate of growth of
a primordial black hole due to accretion of radiation in Brans-Dicke theory 
is considerably smaller than the rate of growth of the cosmological horizon,
thus making available sufficient radiation density for the black hole to
accrete causally. We show that accretion of radiation by Brans-Dicke 
black holes overrides the effect of Hawking evaporation during the radiation
dominated era. The subsequent evaporation of the black holes in later
eras is further modified due to the variable gravitational ``constant'',
and they could survive up to longer times compared to the case of
standard cosmology. We estimate the impact of accretion on modification of
the constraint on their initial mass fraction  obtained
from the $\gamma$-ray background limit from presently evaporating 
primordial black holes. 
\end{abstract}
%\keywords{Brans-Dicke theory, primordial black holes}
\maketitle
%%%%%%%%%%%%%%%%%%%%%%%%%%%%%%%%%%%%%%%%%%%%%%%
\section{Introduction}
%%%%%%%%%%%%%%%%%%%%%%%%%%%%%%%%%%%%%%%%%%%%%%%
The Brans-Dicke (BD) theory \cite{brans and dicke} which was proposed in 1961 
is regarded as a
viable alternative of Einstein's general theory of relativity (GTR). In the BD 
theory, the value of gravitational constant is set by the inverse of a 
time-dependent scalar field which couples to gravity with a coupling 
parameter $\omega$. GTR can be recovered from this BD theory in the limit of 
$\omega \to \infty$. The BD theory has been used in attempts to 
understand many  
cosmological phenomena such as inflation \cite{johri and la}, early and late time behaviour of 
the universe \cite{sahoo}, cosmic acceleration and structure formation \cite{bermar}, and the
coincidence problem\cite{nayak}. It is also well known that BD-type models arise as low
energy effective actions of several higher dimensional Kaluza-Klein and
string theories \cite{asm}. 

Black holes which could be formed in the early Universe through a variety 
of mechanisms are known as primordial black holes [PBHs]. Some of the 
well-studied mechanisms of PBH formation include those due to 
inflation \cite{cgl,kmz}, initial inhomogeneities \cite{carr,swh}, phase transition and critical phenomena in gravitational collapse\cite{khopol,jedam}, bubble collision \cite{kss} or 
the decay of cosmic loops \cite{polzem}  
The formation masses of PBHs could be small enough for them to have evaporated 
completely by the present epoch due to Hawking evaporation \cite{hawk}. 
Early evaporating PBHs 
could account for baryogenesis \cite{bckl,mds} in the universe. On the other
hand, longer lived 
PBHs could act as seeds for structure formation or even as precursors to
supermassive black holes observed presently\cite{mor}. Furthermore, in
certain scenarios it is possible for the PBHs to survive till date and
form a significant component of 
dark matter \cite{blais}. It has been shown
recently that PBHs in the braneworld scenario can efficiently accrete 
radiation \cite{majumdar,gcl,majmuk} 
making them considerably long-lived.  

The possibility of black hole solutions
in BD theory was first proposed by Hawking \cite{hawking}.
Using scalar-tensor gravity theories Barrow and Carr \cite{barrow and carr}  
have studied PBH 
evaporation during various eras.
It has been recently observed that in the context of generalised Brans-Dicke 
theory, inclusion of the
effect of accretion leads to the prolongation of PBH lifetimes \cite{mgs} .
The coexistence
of black holes with a long range scalar field in cosmology could have
interesting consequences \cite{zlo}. The possibility of variation of
fundamental constants of nature over cosmological scales is a fascinating
albeit contentitious issue \cite{davies}, and the black holes themselves
could be used to constrain the variation of fundamental constants \cite{mac}.
Another interesting
issue of gravitational memory of black holes in BD theory has also
been studied \cite{barrow,harada}. 

Accretion of radiation by PBHs in the radiation dominated era of
the early universe has been a much debated issue in standard cosmology.
It is widely held that accretion is ineffective for sufficiently
increasing the mass of a PBH \cite{zeldo}, though later works have
pointed out contrary possibilities in certain cases \cite{hacyan,mds,custodio}.
Recently, it has been realized in the context of the braneworld
scenario that the possibility of enhanced accretion is quite favored
due to the modified PBH geometry, as well as the modified early
high energy era of the universe \cite{majumdar,gcl}. Such a feature of
effective early accretion prolonging the PBH lifetime by significant
orders could also be valid for other modified gravity scenarios,
as has already been shown in the context of a generalized scalar-tensor
model \cite{mgs}.

The motivation for the present work is to study how accretion of radiation
during the radiation dominated era impacts the evolution of primordial
black holes in BD theory. Cosmological solutions during different eras
in the BD theory were obtained by Barrow \cite{jdbarrow}. Here we consider 
together the processes of accretion
of radiation and Hawking evaporation for PBHs in BD theory using the
solutions for the scale factor and the gravitational ``constant'' as
used by Barrow and Carr \cite{barrow and carr}. In the present work we do not consider
the effect of gravitational memory \cite{barrow,harada} on PBH evolution, but
assume that the evolution for the BD field is similar for the PBHs 
as it is for the whole universe. Cosmological observations could be
used to impose constraints on the density of primordial black holes
at various eras \cite{green and carr}. Within this formalism we also estimate 
the impact of accretion in modifying 
the constraint on their initial mass fraction in BD theory obtained
from the $\gamma$-ray background limit from presently evaporating 
primordial black holes.

%%%%%%%%%%%%%%%%%%%%%%%%%%%%%%%%%%%%%%%%%%%%%
\section{PBHs in Brans-Dicke Theory}
%%%%%%%%%%%%%%%%%%%%%%%%%%%%%%%%%%%%%%%%%%%%%
For a spatially flat($k=0$) FRW universe with scale factor \textit{$a$} , the Einstein equations and the equation of motion for the JBD field $\Phi$ take the form
\be \label{bd1} 
\frac{\dot{a}^2}{a^2}+\frac{\dot{a}}{a}\frac{\dot{\Phi}}{\Phi}-\frac{\omega}{6}\frac{\dot{\Phi}^2}{\Phi^2}=\frac{8\pi\rho}{3\Phi}
\ee
\be \label{bd2}
2\frac{\ddot{a}}{a}+\frac{\dot{a}^2}{a^2}+2\frac{\dot{a}}{a}\frac{\dot{\Phi}}{\Phi}+\frac{\omega}{2}\frac{\dot{\Phi}^2}{\Phi^2}+\frac{\ddot{\Phi}}{\Phi}=-\frac{8\pi p}{\Phi}
\ee
\be \label{bd3}
\frac{\ddot{\Phi}}{8\pi}+3\frac{\dot{a}}{a}\frac{\dot{\Phi}}{8\pi}=\frac{\rho-3p}{2\omega+3}  .
\ee
The energy conservation equation is given by
\be \label{ec}
\dot{\rho}+3(\gamma+1)H\rho=0  
\ee
assuming a  perfect fluid equation of state $p=\gamma\rho$ .

Barrow and Carr \cite{barrow and carr} have obtained the following solutions for $a$ and $G$ for different eras, as
\ba 
a(t) \propto \left\{ 
\begin{array}{rr}
t^{(1-\sqrt{n})/3} &  (t<t_1)\\
t^{1/2} &  (t_1<t<t_e)\\
t^{(2-n)/3} &  (t>t_e) 
\end{array}
\right.
\label{sola}
\ea
and
\ba 
G(t)= \left\{
\begin{array}{rr}
G_0\Big(\frac{t_1}{t}\Big)^{\sqrt{n}}\Big(\frac{t_0}{t_e}\Big)^n & (t<t_1)\\
G_0\Big(\frac{t_0}{t_e}\Big)^n & (t_1<t<t_e)\\
G_0\Big(\frac{t_0}{t}\Big)^n & (t>t_e)
\end{array}
\right.
\label{solg}
\ea
where $t_1 \sim$ is the time of starting of radiation dominated era, 
$t_e \sim$ is the era of matter-radiation equality, $t_0 \sim$ is the present 
time, $G_0 \sim$ is the present value of $G$ $\simeq \frac{t_{pl}}{M_{pl}}$,
and $~~n$ is a parameter related to $\omega$, i.e., $n=\frac{2}{4+3\omega}$ .
Since solar system observations \cite{bit} require that $\omega$ be large ($\omega \geq 10^4$), $n$ is very small ($n \leq 0.00007$) .

Barrow and Carr have considered only evaporation of the primodial black holes due to Hawking radiation. If we consider accretion which is effective in radiation dominated era, then the primordial black holes take more time to evaporate. Let
us study how accretion changes the life time of the primodial black holes.
For a primordial black hole immersed in the radiation field, the accretion of radiation leads to the increase of its mass with the rate given by
\be  \label{acc1}
\dot{M}_{acc}=4\pi fR_{BH}^2 \rho_R
\ee
where $R_{BH}=2MG$ is the black hole radius,
$~\rho_R =\frac{3}{8\pi G} \Big(\frac{\dot{a}}{a}\Big)^2$ is the
radiation energy density surrounding the black hole.
and $f \sim$ is the accretion efficiency. The value of the efficiency
of accretion $f$ depends upon complex physical processes such as the
mean free paths of the particles comprising the radiation surrounding
the PBHs. Any peculiar velocity of the PBH with respect to the cosmic
frame could increase the value of $f$ \cite{mds}. Since the precise value of $f$
is unknown, it is customary \cite{gcl} to take the accretion rate
to be proportional to the product of the surface area of the PBH and the
energy density of radiation with $f \sim O(1)$. 
After substituting the above expressions for $R_{BH}$ and $\rho_R$ equation (\ref{acc1}) becomes
\be \label{acc2}
\dot{M}_{acc}=6fG\Big(\frac{\dot{a}}{a}\Big)^2 M^2
\ee

Accretion is effective only in the radiation dominated era. So the primordial black holes which exist during the radiation dominated era are affected by accretion.
Using the solutions for the scale factor $a(t)$ (\ref{sola}) and $G(t)$ (\ref{solg}) in equation (\ref{acc2}), we get for the radiation dominated era
\be \label{acc3}
M(t)=\Big[M_i^{-1} +\frac{3}{2}fG_0\Big(\frac{t_0}{t_e}\Big)^n\Big(\frac{1}{t}-\frac{1}{t_i}\Big)\Big]^{-1}  
\ee
where $M_i$ is the black hole mass at it's formation time $t_i >t_1$. Assuming
the standard mechanism for PBH formation due to gravitational collapse of
density perturbations at the cosmological horizon scale \cite{carr}, we 
have $M_i \simeq G^{-1}(t_i)t_i$ .
Since the horizon mass grows as $M_H(t)\sim G^{-1}t$, one finds that $M_H$ grows faster than the black hole mass $M_{BH}$ which for large times asymptotes to
$M_i(1-3/2f)^{-1}$. Thus enough radiation density is 
available within the cosmological horizon for a PBH to accrete causally, making
accretion effective in this scenario. However, the maximum accretion efficiency
cannot exceed $f = 2/3$, thereby making it improbable for the overall mass of 
a PBH to increase much due to accretion. Nonetheless, the occurrence of
accretion prolongs the onset of the evaporating era for the PBH thereby
prolonging its lifetime considerably which in turn could significantly impact
the observational constraints on PBHs in different eras. 

Due to Hawking evaporation, the rate at which the primordial black hole mass decreases is given by
\be \label{eva1}
\dot{M}_{evap}=-4\pi R_{BH}^2 a_H T_{BH}^4
\ee
where $a_H$ is the Stefan-Boltzmann constant,
and $~~T_{BH} =\frac{1}{8\pi GM}$ is the Hawking temperature.
Now
\be \label{evap}
\dot{M}_{evap}=-\frac{a_H}{256\pi^3} \frac{1}{G^2M^2}  .
\ee

If we consider both evaporation and accretion simultaneously, then the rate at which the primordial black hole mass changes is given by
\be \label{total}
\dot{M}_{BH}=6fG\Big(\frac{\dot{a}}{a}\Big)^2 M^2-\frac{a_H}{256 \pi^3} \frac{1}{G^2M^2}  .
\ee
This equation can not be solved exactly. But we can very well approximate it
during different regimes when either accretion or evaporation is the dominant 
process. Subsequently, we also integrate it by using numerical methods, which
corroborates our approximation.

%%%%%%%%%%%%%%%%%%%%%%%%%%%%%%%%%%%%%%%%%%%%%%%%%%%%%%%
\section{PBH dynamics in different eras}
%%%%%%%%%%%%%%%%%%%%%%%%%%%%%%%%%%%%%%%%%%%%%%%%%%%%%%%
\subsection{ For $t<t_1$ :}
Hawking evaporation rate for this era is given by
\be \label{beva}
\dot{M}_{evap}=-\alpha \Big(\frac{t_e}{t_0}\Big)^{2n} \Big(\frac{1}{t_1}\Big)^{2\sqrt{n}}  \Big(\frac{t^{2\sqrt{n}}}{M^2}\Big) 
\ee
where $\alpha=\frac{a_H}{256 \pi^3 G_0^2}$ .
Integrating the above equation, one gets 
\ba \label{btot}
M^3=M_i^3+3 \alpha \Big(1+2\sqrt{n}\Big)^{-1} \Big(\frac{t_e}{t_0}\Big)^{2n} \Big(t_1\Big)^{-2 \sqrt{n}} \nonumber \\  \Big(t_i^{2\sqrt{n}+1} - t^{2\sqrt{n}+1}\Big)
\ea
This regime corresponds to the BD field dominated dynamics, where the radiation
density is only subdominant. Assuming accretion is not effective in this era, we get same result as that of
Barrow and Carr \cite{barrow and carr} for the evaporation time ($M=0$) . 
\ba
\tau=\Big[(3\alpha)^{-1}(1+2\sqrt{n})\Big(\frac{t_0}{t_e}\Big)^{2n}M_i^3t_1^{2\sqrt{n}}+t_i^{1+2\sqrt{n}}\Big]^{\frac{1}{1+2\sqrt{n}}} \nonumber \\
\label{lt}
\ea

\subsection{For $t_1<t<t_e$ :}
This period corresponds to the radiation dominated era. Here we consider two possibilities: PBHs created before $t_1$ and those created after $t_1$.\\

\textbf{CASE-I} ($t_i<t_1$) :\\
The PBH evaporation equation (\ref{evap}) becomes
\ba \label{eva1}
\int_{M_i}^M M^2 \,dM=-\alpha\Big[\int_{t_i}^{t_1} {{\Big(\frac{1}{t_1}\Big)}^{2\sqrt{n}} {\Big(\frac{t_e}{t_0}\Big)}^{2n}  t^{2\sqrt{n}}} \,dt \nonumber \\ + \int_{t_1}^t {{\Big(\frac{t_e}{t_0}\Big)}^{2n}} \,dt\Big]  .
\ea
Integrating this equation, one gets
\ba
M^3=M_i^3+3\alpha \Big(\frac{t_e}{t_0}\Big)^{2n}\Big[t_1 - t \nonumber\\
+ \Big(\frac{1}{t_1}\Big)^{2\sqrt{n}}{\Big(1+2\sqrt{n}\Big)}^{-1} 
\Big( t_i^{2\sqrt{n}+1}- t_1^{2\sqrt{n}+1}\Big)\Big] .  
\label{reva1}
\ea

Now considering both evaporation and accretion, we obtain
\ba
\dot{M}_{BH}= -\alpha \Big(\frac{1}{t_1}\Big)^{2\sqrt{n}}\Big(\frac{t_e}{t_0}\Big)^{2n}\Big(\frac{t^{2\sqrt{n}}}{M^2}\Big)
\label{tot1}
\ea
during the BD field dominated era ($t < t_1$)
and 
\ba
\dot{M}_{BH} = \frac{3}{2}fG_0\Big(\frac{t_0}{t_e}\Big)^n\Big(\frac{M^2}{t^2}\Big)-\alpha \Big(\frac{t_e}{t_0}\Big)^{2n}\frac{1}{M^2}
\label{tot11}
\ea
for $t >> t_1$ in the radiation dominated era.
Since accretion is effective only during the radiation dominated era, 
Eq.(\ref{tot1}) is first integrated over the time period $t_i$ to $t_1$ and 
then Eq.(\ref{tot11}) is then integrated over the time period $t_1$ to $t$
with the initial condition $M_{BH}(t_1)$ obtained from the solution of
Eq.(\ref{tot1}).
The results  of numerical integration of the above equation with 
several values of $M_i$ are presented in the figure-1. We assume that radiation 
domination sets in at the GUT scale, i.e., $t_1=10^{-35}s$. We find that the
PBHs formed before the onset of radiation domination, i.e., for $t_i < t_1$, 
evaporate out completely during  the radiation dominated era. Figure 1 shows
the variation of PBH masses for three different values of $M_i$ with 
$M_i \sim 990 g$ corresponding to the formation time $t \sim t_1$. Though here the accretion efficiency $f=1/3$ , it can be verified that even for 
larger $f$, accretion in this case is not able to prolong PBH lifetimes
beyond the radiation dominated era, i.e., those  PBHs which form at a time
$t_i < t_1$, have evaporations times $t_{evap} < t_e$. Note that though in
our analysis we assume radiation domination at the GUT scale, it is
possible for the radiation domination to set in much 
later, i.e., $t_1>>10^{-35}s$ in certain scenarios, e.g., gravitino production
could constrain the reheating temperature to be much lower \cite{rubakov}.
In such a case of low reheating temperature, accretion of radiation will
be effective for a shorter window of time corresponding to a shortened span
of radiation domination, and the evolution of PBHs would be closer to
the case in BD theory without accretion \cite{barrow and carr}.

\vskip 0.1in

\begin{figure}[h]
\includegraphics[scale=0.8]{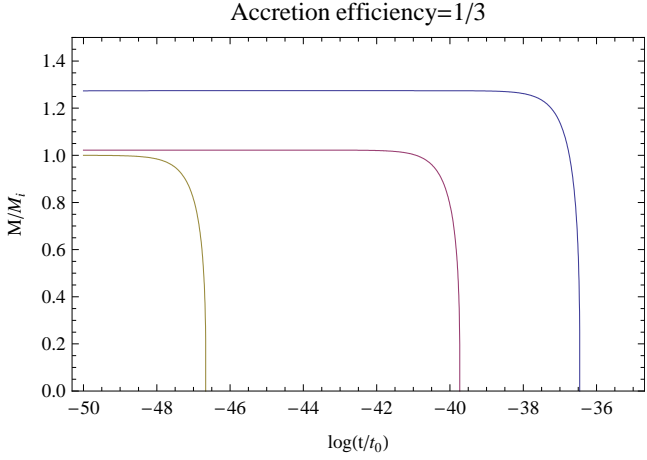}
\caption{Evolution of PBH masses with time ($t_0$ being the present age
of the universe) for different initial mass $M_i=0.5g, 100g, 990g$, but
with same accretion efficiency $f=1/3$.}
\label{fig1}
\end{figure}

\textbf{CASE-II}  ($t_i>t_1$) :\\

Taking both accretion and evaporation into account, we can write
\be \label{tot2}
\dot{M}_{BH}=\frac{3}{2}fG_0\Big(\frac{t_0}{t_e}\Big)^n\Big(\frac{M^2}{t^2}\Big)-\alpha\Big(\frac{t_e}{t_0}\Big)^{2n}\frac{1}{M^2} .
\ee
For PBHs with formation mass $M_i^2 > \frac{a_HG^{-1}}{384 f}$, the 
magnitude of the first term (accretion) exceeds that of the second term 
(evaporation).
In the radiation dominated era for a PBH whose formation mass satisfies
the above relation, accretion is dominant upto a value of $t$, say $t_{c}$ at which accretion equals evaporation (the PBH mass
rises to a maximum value $M_{max}$ at this stage), and after that evaporation dominates over accretion. 
For $t=t_{c}$, the magnitude of the accretion term is equal to the magnitude of evaporation term. So for the radiation dominated era, equation (\ref{tot2}) implies,
\be \label{equal}
\frac{3}{2}fG_0\Big(\frac{t_0}{t_e}\Big)^n\Big(\frac{M_{max}^2}{t^2}\Big)=\alpha\Big(\frac{t_e}{t_0}\Big)^{2n}\Big(\frac{1}{M_{max}^2}\Big)  
\ee
which gives
\be \label{max1}
M_{max}=\Big(\frac{A}{f}\Big)^{\frac{1}{4}} \times \Big(t_{c}\Big)^{\frac{1}{2}}
\ee
where $A=\frac{2}{3}G_0^{-1}\alpha\Big(\frac{t_e}{t_0}\Big)^{3n}$ and $M_{max}=M(t_c)$ .
But from the PBH accretion equation (\ref{acc3}), we have
\be \label{max2}
M_{max}=M_i\Big[1+\frac{3}{2}f\Big(\frac{t_i}{t_c}-1\Big)\Big]^{-1}.
\ee
Equating above two expressions for $M_{max}$, one gets
\be \label{crit}
t_c^{1/2}=\Big(\frac{f}{A}\Big)^{\frac{1}{4}} \times \frac{M_i}{1-\frac{3}{2}f} 
\ee
and
\be \label{max3}
M_{max}=\frac{M_i}{1-\frac{3}{2}f}  
\ee
 which again stipulates that $f<\frac{2}{3}$.

Considering evaporation  from $t_c$ onwards, we get
\be \label{eqm}
M=M_{max}\Big[1+3\alpha\Big(\frac{t_e}{t_0}\Big)^{2n}\Big(\frac{t_{c}}{M_{max}^3}\Big)\Big\{1-\Big(\frac{t}{t_{c}}\Big)\Big\}\Big]^{\frac{1}{3}}.
\ee
So the evaporation time for these PBHs are given by
\be \label{ceva}
t_{evap}=t_{c}\Big[1+\Big(3\alpha\Big)^{-1}\Big(\frac{t_0}{t_e}\Big)^{2n}\Big(\frac{M_{max}^{3}}{t_{c}}\Big)\Big].
\ee

\subsection{For $t>t_e$ :}

The PBHs which are formed before radiation domination completely evaporate
out during the radiation dominated era. So for $t>t_e$, only those PBHs  which 
are formed after $t_1$ exist. The
PBH evaporation equation (\ref{evap}) 
can be written as
\be \label{eva3}
\int_{M_i}^{M} M^2 \,dM=-\alpha\Big[\int_{t_i}^{t_e} \Big(\frac{t_e}{t_0}\Big)^{2n} \,dt+\int_{t_e}^{t} {(t_0)}^{-2n}t^{2n} \,dt\Big]  .
\ee

Taking both accretion and evaporation into account, one gets
\ba
\dot{M}_{BH}= \frac{3}{2}fG_0\Big(\frac{t_0}{t_e}\Big)^n\Big(\frac{M^2}{t^2}\Big)-\alpha\Big(\frac{t_e}{t_0}\Big)^{2n}\frac{1}{M^2}
\label{ltot}
\ea
during the radiation dominated era
and
\ba
\dot{M}_{BH}=  -\alpha\Big(\frac{1}{t_0}\Big)^{2n} \Big(\frac{t^{2n}}{M^2}\Big)
\label{ltot11}
\ea
during the matter dominated era.
In order to obtain the PBH lifetime, one can numerically integrate the
above equations (\ref{ltot}) and (\ref{ltot11}). 
Eq.(\ref{ltot}) corresponding to accretion and evaporation 
 during the radiation dominated era is integrated over the 
period $t_i$ to $t_e$ and the value of $M_{BH}(t_e)$ is obtained and used as
the initial condition for integrating Eq.(\ref{ltot11})  
over the period $t_e$ to $t$.
The results  of numerical integration are displayed for a particular initial 
mass in figure-2. (In the figure we do not show the early part of their
evolution where their mass increases by a bit due to accretion and then 
stays nearly constant for a long period of time). 
One sees that depending on the accretion efficiency $f$,
the lifetimes of the PBHs formed during the radiation dominated era
could exceed the present age of the universe $t_0$.

\begin{figure}[h]
\includegraphics[scale=0.8]{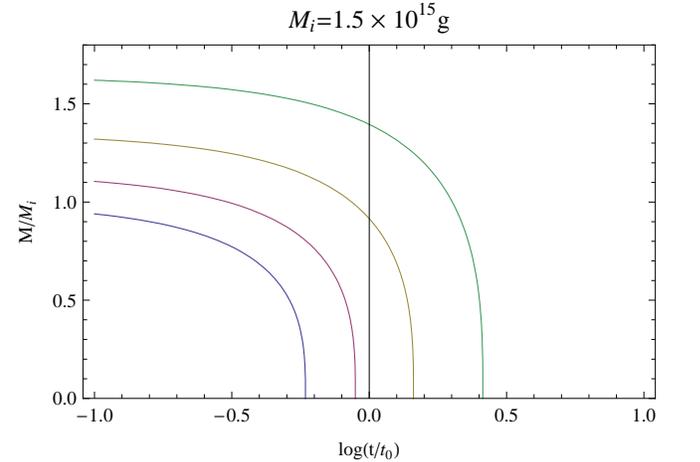}
\caption{The late time evolution of PBH masses (with the same initial mass 
$M_i = 1.5 \times 10^{15}g$) for various accretion efficiency values 
$f=0, 0.2, 0.4, 0.6$.}
\label{fig2}
\end{figure}

On the other hand, integrating the last term in equation (\ref{ltot}) one can
obtain the mass of the PBHs which survive beyond the radiation dominated era,
given by 
\ba
M=M_e\Big[1+3\alpha\Big(2n+1\Big)^{-1}\Big(\frac{t_e}{t_0}\Big)^{2n}\Big(\frac{t_e}{M_e^3}\Big)\nonumber \\ \Big\{1-\Big(\frac{t}{t_e}\Big)^{2n+1}\Big\}\Big]^{\frac{1}{3}}
\label{mase}
\ea
where $M_e \equiv M(t_e)$ is obtained by integration of the PBH accretion 
equation (\ref{acc2}) over the period $t_i$ to $t_e$ to be
\be \label{lacc}
M_e=M_i\Big[1+\frac{3}{2}fG_0M_i\Big(\frac{t_0}{t_e}\Big)^n\Big(\frac{1}{t_i}-\frac{1}{t_e}\Big)\Big].
\ee
Hence, the PBH lifetime is given by
\be \label{tev}
t_{evap}=t_e\Big[1+\Big(3\alpha\Big)^{-1}(2n+1)\Big(\frac{t_0}{t_e}\Big)^{2n}\Big(\frac{M_e^3}{t_e}\Big)\Big]^{\frac{1}{2n+1}}.
\ee
Further, from equation (\ref{eqm}), we have
\ba
M_e=M_{max}\Big[1+3\alpha\Big(\frac{t_e}{t_0}\Big)^{2n}\Big(\frac{t_c}{M_{max}^3}\Big)\nonumber \\ \Big\{1-\Big(\frac{t_e}{t_c}\Big)\Big\}\Big]^{\frac{1}{3}}.
\label{fmas}
\ea
Now using the equations (\ref{tev}) and  (\ref{max3}), we get 
\ba
M_i \approx \Big\{1-\frac{3}{2}f\Big\} \times \Big[3\alpha\Big(\frac{t_e}{t_0}\Big)^{2n}t_e\Big\{1+\Big(2n+1\Big)^{-1}\nonumber \\ 
\Big\{\Big(\frac{t_{evap}}{t_e}\Big)^{2n+1}-1\Big\}\Big\}\Big]^{\frac{1}{3}}  .
\label{fm}
\ea
This enables us to invert the PBH lifetime relation in order to obtain the
formation time for a PBH given its time of evaporation $t_{evap}$, 
\ba
t_i \approx G_0\Big\{1-\frac{3}{2}f\Big\}\Big(\frac{t_{0}}{t_e}\Big)^n \times \Big[3\alpha\Big(\frac{t_e}{t_{0}}\Big)^{2n}t_e \nonumber \\ \Big\{1+\Big(2n+1\Big)^{-1}\Big(\frac{t_{evap}}{t_e}\Big)^{2n+1}\Big\}\Big]^{\frac{1}{3}}  .
\label{ft}
\ea
The above expression is useful for the purpose of evaluating the constraints
on the initial mass or formation time of the PBHs in terms of the observational
constraints on evaporating black holes at particular eras
in this Brans-Dicke cosmology.
For the present we compute as examples the initial masses of the PBHS for
two cases: (i) PBHs that
are evaporating at the present era, and (ii) the PBHs that will evaporate 
when the universe is ten times older than its present age. These values are
computed using the analytical result (\ref{ft}) and displayed in the Table~I.
We have also computed $t_i$ and $M_i$ for different values of the
accretion efficiency $f$ from numerical integration of equation (\ref{ltot}).
We find that our analytical approach gives results that agree up to
three decimal places with the numerical results, thus validating 
the division of the evolution of a PBH into two distinct eras dominated
by accretion and evaporation dynamics respectively, that we have done.

\begin{table}
%$~~~~~~~~~~~~~~~~~$Table-1\\
\begin{tabular}[c]{|c|c|c|c|c|}
\hline
\multicolumn{5}{|c|}{$t_{evap}=t_0$ \hskip 0.6in $t_{evap}=10 \times t_0$}\\
\hline
$f$  &  $t_i \times 10^{-23}$s &  $M_i \times 10^{15}$g  &  $t_i\times 10^{-23}$s  &  $M_i\times 10^{15}$g\\
\hline
$0$  &  $2.369$   &  $2.366 $ & $5.105$ & $5.099$\\
\hline
$0.2$  &  $1.658$  &  $1.656$ & $3.573$ & $3.569$\\
\hline
$0.4$  &  $0.947$  &  $0.946$ & $2.042$ & $2.039$\\
\hline
$0.6$  &  $0.236$  &  $0.236$ & $0.510$ & $0.509$\\
\hline
\end{tabular}
\caption{The formation times and initial masses corresponding to
two specific evaporating eras of the PBHs are displayed for several
accretion efficiencies.}
\end{table}

%%%%%%%%%%%%%%%%%%%%%%%%%%%%%%%%%%%%%%%%%%%%%
\section{Constraints on the PBH mass fraction}
%%%%%%%%%%%%%%%%%%%%%%%%%%%%%%%%%%%%%%%%%%%%%

Surviving PBHs at any cosmological era contribute to the matter density of 
the universe at that era. Further, PBHs impact different processes by the end 
products of their Hawking evaporation. Various cosmological observations
can be used to impose constraints on the number density of black
holes present during different cosmological eras. These constraints 
can in turn be
used for imposing limits of the initial mass spectrum of PBHs in various
formation mechanisms pertaining to different cosmological models. In standard 
cosmology, a variety of constraints such as coming from considerations of 
overall
density, nuleosynthesis, entropy, distortions of the CMBR spectrum, and
stable relics, have been 
obtained \cite{green and carr}. It has been observed that a particular
stringent set of constraints arise from the 
limits of the $\gamma$-ray background
\cite{mac and carr}, and also independently 
from the observed galactic anti-protons and antideuterons
\cite{barrau}. In the present analysis we will just focus on the $\gamma$-ray
background limit in order to obtain bounds on the initial mass spectrum
of PBHs in BD theory with accretion.

The fraction of the Universe's mass going into PBHs at time $t$ is
given by \cite{carr} 
\be \label{bet1}
\beta(t)=\Big[\frac{\Omega_{PBH}(t)}{\Omega_R}\Big](1+z)^{-1}
\ee
where $\Omega_{PBH}(t)$ is the present density parameter associated with PBHs 
forming at time $t$, $z$ is the redshift associated with time $t$ and $\Omega_R$ is the present microwave background density having value $10^{-4}$.
Observations of the $\gamma$-ray background, as well as those of the 
antiprotons  from galactic sources impose bounds on the present cosmological
PBH density given by \cite{barrow and carr,mac and carr,barrau}
\be
\Omega_{PBH}(t) < 10^{-8}
\label{densitybound}
\ee
Let us here consider the PBHs that are formed during the radiation
dominated era, i.e., $t_1<t_i<t_e$. For them, one has
\be
(1+z)^{-1}=\Big(\frac{t}{t_e}\Big)^{\frac{1}{2}} \Big(\frac{t_e}{t_0}\Big)^{\frac{2-n}{3}}.
\label{redshift}
\ee
Using equations ({\ref{densitybound}) and (\ref{redshift}) in 
equation (\ref{bet1}), one gets a bound on the density fraction given by
\be \label{bet3}
\beta(t) <  10^{-4} \times \Big(\frac{t}{t_e}\Big)^{\frac{1}{2}} \times \Big(\frac{t_e}{t_0}\Big)^{\frac{2-n}{3}}  .
\ee
The above bound pertains to those PBHs which are evaporating at the present era.
Again, for $t_1<t <t_e $, one has $M=G^{-1}t=G_0^{-1}t\Big(\frac{t_e}{t_0}\Big)^n$ . 
Thus, the fraction of the Universe going into PBHs with formation mass $M_i$ is 
\be \label{bet4}
\beta(M_i) < 10^{-4} \times \Big(\frac{M_i}{M_e}\Big)^{\frac{1}{2}} \times \Big(\frac{t_e}{t_0}\Big)^{\frac{2-n}{3}}  .
\ee
We can now use the expressions for $M_i$ (\ref{fm}) in terms of the evaporation
time $t_{evap} = t_0$ to obtain the values of $\beta(M_i)$ corresponding
to various values of the accretion efficiency $f$. These are displayed in
Table~II. It was shown earlier \cite{mac and carr} how the standard constraints
on the initial mass spectrum are modified in BD theory without accretion. Here
we observe that increase of accretion efficiency makes the 
limits on the initial mass fraction more stringent. It may be noted that 
similar considerations would also apply to constraints on the initial
mass fraction $\beta(M_i)$ obtained from other physical considerations
such as those due to entropy or nucleosynthesis bounds. The relevant values
for the initial PBH masses $M_i$ corresponding to the PBHs evaporating
earlier to impact entropy production or nucleosynthesis would
of course be much lower than the values of $M_i$ for which we have
applied the $\gamma$-ray bounds, since these latter PBHs are those that
are evaporating in the present era.  The BD dynamics alters
the evaporation rate for the PBHs thus loosening somewhat the bounds on
$\beta(M_i)$ as shown by Barrow and Carr \cite{barrow and carr}. However,
inclusion of accretion reverses the scenario since accretion is more
effective for a longer duration for PBHs with smaller $M_i$ which have a 
chance to grow more. As a consequence, $\Omega_{PBH}(t)$ increases, and
the standard constraints on $\beta(M_i)$ due to nucleosynthesis, entropy,
etc., are tightened further in the BD
scenario with accretion.

%$~~~~~~~~~~~~~~~~~$Table-5\\
\begin{table}
\begin{tabular}[c]{|c|c|c|}
\hline
\multicolumn{3}{|c|}{$t_{evap}=t_0$}\\
\hline
$f$  &  $M_i \times 10^{15}$g  &  $\beta(M_i) <$\\
\hline
$0$  &  $2.366$   &  $5.71 \times 10^{-26}$\\
\hline
$1/6$  &  $1.775$  &  $4.95 \times 10^{-26}$\\
\hline
$1/3$  &  $1.183$  &  $4.03 \times 10^{-26}$\\
\hline
$1/2$  &  $0.592$  &  $2.85 \times 10^{-26}$\\
\hline
$3/5$  &  $0.236$  &  $1.81 \times 10^{-26}$\\
\hline
\end{tabular}
\caption{Upper bounds on the initial mass fraction of PBHs that are
evaporating today for various accretion efficiencies $f$.}
\end{table}

\section{Conclusions}

In this paper we have considered the evolution of primordial black holes
in Brans-Dicke theory. We use the framework of a particular cosmological
solution \cite{jdbarrow} where the gravitational ``constant'' $G$
can have a much larger value in the early universe compared to its present
strength. We show that accretion of radiation during the radiation dominated
era is an effective process in this theory,  overriding the
mass loss of the PBHs due to Hawking evaporation. The cosmological horizon
for  a PBH grows at a faster rate compared to the rate of growth of the PBH
due to accretion, thus making enough radiation available for the PBH
to accrete during this stage. Though the net gain of mass
through accretion is not significant,  it postpones the time for
evaporation to take over once accretion ceases to be effective at the
onset of the matter dominated era. The evaporation rate for a BD PBH
is itself modifed compared to that in standard cosmology due the variable
$G$. We show that the lifetime for PBHs could be enhanced depending upon
the accretion efficiency $f$, making the PBHs that are supposed to be
evaporating now ($t_0$) live longer by a factor of 
$[\frac{1}{1-\frac{3}{2}f}]^{3/(2n+1)} \times t_0$.

The cosmological evolution of PBHs could lead to various interesting
consequences during different eras. The PBHs with smaller masses that
are formed during the BD field dominated very early era evaporate out
completely during the radiation dominated era. However, those PBHs
that are formed later during the radiation dominated era survive much
longer. These are the ones that through their
evaporation could impact various cosmological
processes such as nucleosynthesis and photon decoupling. There exists a 
variety of observational constraints on the PBHs in standard cosmology 
\cite{green and carr}.
Within the context of BD theory Barrow and Carr \cite{barrow and carr}
have evaluated the impact of density bounds on the PBHs evaporating
today on their initial mass spectrum. Here we use the observational
limits on the $\gamma$-ray background \cite{mac and carr} to compute the
effect of accretion on constraining the initial mass fraction of the PBHs.  
The departure of these constraints from those in standard cosmology
are quite sensitive to the accretion efficiency.

Finally, it may be noted that there exist other interesting
cosmological solutions for the Brans-Dicke theory and its extensions
to more general scalar-tensor models \cite{barrow2}. In a recent paper
one such solution \cite{sahoo} with a time-evolving BD coupling 
parameter $\omega$  
was used to study cosmological PBH evolution \cite{mgs}. 
It could be worthwhile to
remap the standard observational constraints on PBHs in
such models in order to estimate the impact of rapidly varying $G$, as
well as possible effective accretion on the constraints. Another important issue
in PBH evolution could be the impact of the back reaction of the PBHs
on the local background value of the BD field resulting from the local
change of the density due to the PBHs.  Back reaction could indeed
lead to non-trivial consequences on cosmological evolution \cite{buchert},
and in the present context it might be interesting to see in what way any
resulting modification could in turn impact the evolution of black
holes in Brans-Dicke theory.

\section*{Acknowledgements}
A.S.M. acknowledges support form the DST project SR/S2/PU-16/2007.
B. Nayak is thankful to S. N. Bose National Centre for Basic Sciences, Kolkata, India, for providing hospitality with computational facility. He would also like to thank, the Council of Scientific and Industrial Research, Government of India, for the award of JRF, F.No. $09/173(0125)/2007-EMR-I$ . 

%%%%%%%%%%%%%%%%%%%%%%%%%%%%%%%%%%%%%%%%%%%%%%%%%%%%%%%%%%%%%%%

\end{document}